# Broad nonlocal spectrum in the Pb-InSb hybrid three terminals for potential realization of Kitaev chains

Guoan Li,[1,5,*] Xiaofan Shi,[1,5,*] Ruixuan Zhang,[1,5,*] Yuxiao Song,[1,5,*] Marco Rossi,[2] Ghada Badawy,[2] Zhiyuan Zhang,[1,5] Anqi Wang,[1] Xingchen Guo,[1,5] Xiao Deng,[1,5] Xiao Chen,[1,5] Liangqian Xu,[1,5] Bingbing Tong,[1] Peiling Li,[1] Xiaohui Song,[1] Zhaozheng Lyu,[1] Guangtong Liu,[1,6] Fanming Qu,[1,5,6] Michał P. Nowak,[3] Paweł Wójcik,[4] Ziwei Dou,[1,†] Erik P. A. M. Bakkers,[2,‡] Li Lu,[1,5,6,§] and Jie Shen[1,∥]

[1]Beijing National Laboratory for Condensed Matter Physics, Institute of Physics, Chinese Academy of Sciences, Beijing 100190, China
[2]Department of Applied Physics, Eindhoven University of Technology, 5700 MB Eindhoven, The Netherlands
[3]AGH University of Krakow, Academic Centre for Materials and Nanotechnology, al. A. Mickiewicza 30, 30-059 Krakow, Poland
[4]AGH University of Krakow, Faculty of Physics and Applied Computer Science, al. A. Mickiewicza 30, 30-059 Krakow, Poland
[5]School of Physical Sciences, University of Chinese Academy of Sciences, Beijing 100049, China
[6]Hefei National Laboratory, Hefei 230088, China

*These authors contributed equally to this work.
†Contact author: ziweidou@iphy.ac.cn
‡Contact author: e.p.a.m.bakkers@tue.nl
§Contact author: lilu@iphy.ac.cn
∥Contact author: shenjie@iphy.ac.cn

**ABSTRACT:** Hybrid superconductor–semiconductor (SC-SM) nanowires remain one of the foremost platforms for engineering topological superconductivity and Majorana zero modes (MZMs) towards fault-tolerant topological qubits, especially with the rapid development of artificial Kitaev chains. In contrast to the widely used aluminum (Al)-based hybrids, lead (Pb) offers a bulk superconducting gap of ≈ 1.4 meV and a critical temperature of ≈ 7.2 K, giving rise to a proximity-induced gap that is roughly five times larger than that obtained with Al. Here we present the first three-terminal Pb-hybrid devices and perform non-local differential-conductance spectroscopy on this platform. The non-local measurement simultaneously resolves a dual-gap feature of the parent Pb gap and the large, hard, gate-tunable induced superconducting gap, distinguished by a switch between electron- and hole-like dissipation processes. Within the induced gap we observe several types of Andreev bound states (ABSs) that undergo singlet-doublet transitions. Moreover, by tuning gate voltages we achieve gate-controlled resonating sign reversals of the non-local conductance, identifying three distinct regimes that correspond to different configurations of quantum-dot (QD) resonances (single-resonance, double-resonance, and series-resonance). Finally, the coupling between ABSs and QDs also present and can be modulated from the weak- to strong-coupling limit, indicating the feasibility of realizing the artificial Kitaev chains. Crucially, the robust non-local signatures persist up to temperatures (~1 K) far above the operating temperature of Al-based devices thanks to the unusually large induced gap, thereby widening the accessible parameter space greatly and underscoring the suitability of Pb-based hybrids for implementing warm-temperature artificial Kitaev chains and the topological quantum devices protected by a substantially larger topological gap.

Hybrid superconductor (SC)-semiconductor (SM) systems exploit the proximity effect to combine complementary properties of both materials, including particle-hole symmetry within the Bardeen-Cooper-Schrieffer (BCS) gap, tunability of chemical potential and electron modes, strong spin-orbit coupling (SOC) and effective Landé $g$-factors [1-8]. These features enable the realization of topological phase transitions and the emergence of Majorana zero modes (MZMs) in both Lutchyn-Oreg-type nanowires [2,9,10] and the artificial Kitaev chain architectures [11-15]. While aluminum (Al)-based superconductors have been widely adopted in such systems, their intrinsic limitations—including a small superconducting gap (~ 0.2 meV) [16-19] and weak SOC [4-5]—not only constrain the robustness of Majorana devices but also limit critical topological parameter regimes [20]. For example, Microsoft has reported a potential topological gap on the Al-InAs hybrid nanowires whose size is only 30 μeV, providing very limited protection against thermal fluctuations and material imperfections such as disorder-induced perturbations [20]. Consequently, researchers have sought alternative superconductors such as tin (Sn) and lead (Pb) [5,21]. Pb, a Type-I superconductor with a critical temperature of ~ 7.2 K and a bulk superconducting gap of ~ 1.4 meV [16], which are six-folder larger than that of Al [17-19], uniquely combines high-quality superconducting properties with intrinsic strong SOC [22,23]. Experimental studies by local measurement have demonstrated that Pb induces robust superconducting gaps exceeding 1 meV in InSb nanowires through proximity coupling, accompanied by significant SOC and amplified effective $g$-factors [24,25]—All are indispensable for achieving topological phases hosting MZMs. These capabilities position Pb-nanowire heterostructures as a promising platform for realizing scalable topological quantum systems.

Recent advances emphasize nonlocal measurements in three-terminal devices as a definitive experimental tool for verifying MZMs and topological phase signatures [11,20,26]. Unlike local conductance measurements, nonlocal approaches probe the correlations between spatially separated states, enabling direct observation of zero-bias signatures for the paired MZMs from both ends of hybrid nanowires and distinguishing extended bulk topological phase transitions from trivial local Andreev bound states (ABSs) [27-29]. Furthermore, in artificial Kitaev chain implementations, nonlocal coupling of two side spin-polarized quantum dots by the middle ABSs through elastic co-tunneling (ECT) and crossed Andreev reflection (CAR) directly manifests poor man's MZMs and gain a rapid development recently [11-15,26,30]. This artificial Kitaev chain has gradually emerged as one of the most direct routes to topological qubits because it reproduces the p-wave pairing mechanism of the Kitaev model in a controllable solid-state architecture. However, the Kitaev chain has never been realized on Pb-based hybrid devices. To realize a Kitaev chain, nonlocal measurement and operation in three terminals, including how the in-gap states couple with the side quantum dots (QDs), are essential. These factors underscore the urgent need for systematic nonlocal characterization on Pb-based hybrid three terminals, in particular revealing how to systematically change the global coupling between SC-SM and the resulting extended topological phase transition, as well as the ABS-QDs coupling towards the realization of artificial Kitaev chain. Such studies may rigorously assess their capacity to host MZMs and establish their potential for scalable topological qubits for hybrid Pb-nanowire hybrid system, with the benefit of simultaneously validating large superconducting gaps, strong SOC, and enhanced effective *g*-factors.

*These authors contributed equally to this work.
†Contact author: ziweidou@iphy.ac.cn
‡Contact author: e.p.a.m.bakkers@tue.nl
§Contact author: lilu@iphy.ac.cn
‖Contact author: shenjie@iphy.ac.cn

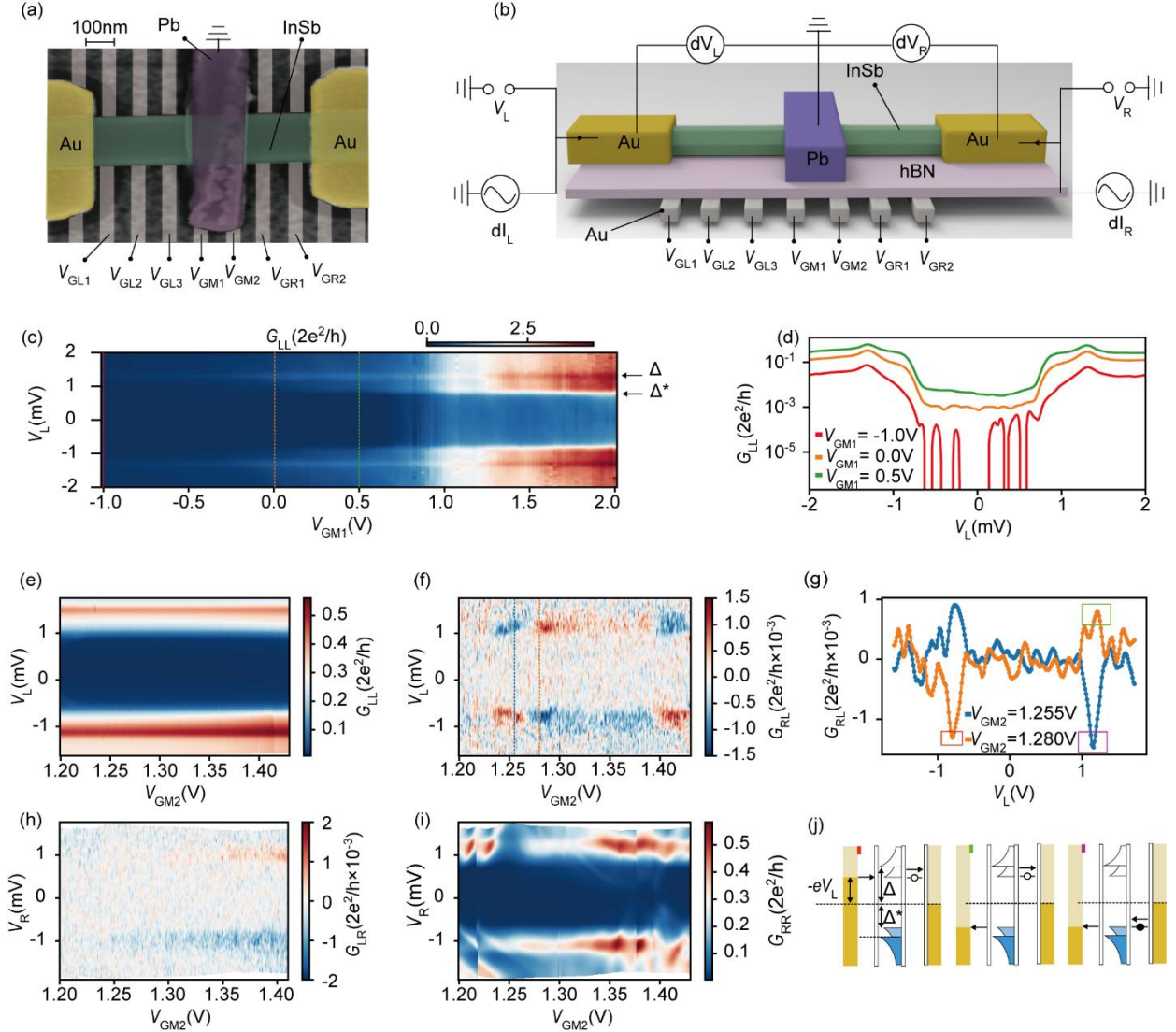

FIG. 1. Large and hard induced gap in InSb-Pb three-terminal devices probed by local and nonlocal transport measurements: (a) False-color SEM image. Pb electrodes (purple) and Ti/Au electrodes (yellow) contact the InSb nanowire (green). Scale bar: 100 nm. (b) Schematics of the measurement setup. See text for the principle and the definition of the full conductance matrix measurements $G_{LL}$, $G_{RL}$, $G_{LR}$, and $G_{RR}$. (c) Local measurement of the left tunnel junction conductance $G_{LL}$ versus $V_L$ and $V_{GM1}$, showing the gap of Pb $\Delta$ and the induced gap $\Delta^*$. (d) Linecuts in (c) showing at least two magnitudes of $G_{LL}$ decreasing inside the gap, demonstrating high hardness. (e,f,h,i) The full conductance matrix measurements versus $V_{GM2}$ and $V_{L,R}$ (for upper and lower panels, respectively). (g) Linecuts of the upper panel of (f) showing dual-antisymmetry (see text) for both $V_L$ and $V_{GM2}$. (j) Illustration of the antisymmetric $G_{RL}$ with $V_L$ indicates the nonlocal current going through the electron or hole states [32]. The gate-modulated sign reversal of $G_{RL}$ (also see (f)) is associated with the junction on the right (nonlocal) side supporting varying electron or hole-like carrier [32]. All the arrows here point to the direction of electron.

*These authors contributed equally to this work.
†Contact author: ziweidou@iphy.ac.cn
‡Contact author: e.p.a.m.bakkers@tue.nl
§Contact author: lilu@iphy.ac.cn
∥Contact author: shenjie@iphy.ac.cn

Here, we fabricated the three terminal devices on Pb-InSb nanowire and perform the first-reported nonlocal measurement for Pb-based hybrid devices. The data clearly reveal the coexistence of the parent Pb gap and the proximity-induced superconducting gap, manifested as the dual-gap structure in the local spectroscopy and as a positive/negative-sign-changing nonlocal differential conductance at finite bias. The induced gap is also proved to be very hard by both the highly suppressed in-gap states in local spectroscopy and the absence of nonlocal signal below the gap. Importantly, the extracted magnitude of the induced gap is almost five times of the previous Al-based devices [11,12,15], allowing the distinguishable controllability of ABSs from singlet to doublet phase transition [31]. Moreover, by inducing quantum dot resonances, we find the positive-to-negative resonating nonlocal switches following the transition between electron- or hole-like resonance properties with three different situations: The first is similar to Al-hybrid device [30,32], with only one single resonance within the superconducting gap: the second is a new phenomenon enabled by the large induced gap, in which two dot resonances coexist inside the gap; the third arises due to a large and another small QD connecting in series. Charge stability diagrams further show that the coupling between ABSs and QDs can be tuned continuously from weak to strong coupling, displaying clear crossing and anti-crossing features that are essential for constructing an artificial Kitaev chain. Our results demonstrate in a Pb-based hybrid, that the three pillars of a Kitaev chain—hard large gap, gate-controlled inter-site coupling, and reliable non-local read-out—can be realized simultaneously. Finally, the nonlocal signatures arising from the large hard induced gap could remain robust up to above 1 K—well above the operating temperature of Al-based hybrids—thereby extending the available parameter space for topological phases and offering a viable route toward "warm" Kitaev chain devices, as well as better scalability with protection from a greatly enhanced topological gap.

The device image and measurement configuration are illustrated in Figs. 1(a,b). The central electrode is a superconducting lead (Pb) electrode, while the top and bottom electrodes are normal metal electrodes composed of titanium and gold. To form quantum dots in both junction regions besides the Pb electrode, we maintained a relatively large distance (200-300 nm) between the superconducting and normal metal electrodes, allowing sufficient gating capability beneath the uncovered nanowire to finely tune its chemical potential via local back gates (see [33] for detailed information on device fabrications). Based on their relative positions to the central hybrid segment—namely left (L), middle (M), and right (R)—we categorized the seven localized back-gate voltages into three groups. For clarity, we further assigned numerical identifiers to each category from left to right. They are thus labeled as: 1. $V_{GL1,2,3}$, 2. $V_{GM1,2}$, and 3. $V_{GR1,2}$

We adopted a full conductance matrix measurement configuration similar to that used in recent experiments achieving three terminal local/nonlocal measurement and minimal Kitaev chains in Al-based devices [11,26,30,34]. The central superconducting electrode is grounded, while DC voltages $V_L$ and $V_R$ are applied to the left and right normal electrodes, respectively. The corresponding DC currents $I_L$ and $I_R$ are defined as positive when flowing from the normal electrodes toward the superconducting electrode. Unless specified otherwise, when varying one terminal's DC voltage, the other remains set to zero. Additionally, we applied an AC signal superimposed on the DC bias using a lock-in amplifier to facilitate full conductance matrix measurements as shown in Eq.1 (see [33] for detailed information on the measurement setups).

$$\begin{pmatrix} G_{LL} & G_{LR} \\ G_{RL} & G_{RR} \end{pmatrix} = \begin{pmatrix} \frac{dI_L}{dV_L} & \frac{dI_L}{dV_R} \\ \frac{dI_R}{dV_L} & \frac{dI_R}{dV_R} \end{pmatrix} \quad (1)$$

First, to verify the presence of a proximitized gap we perform tunneling spectroscopy measurements on the central hybrid section by setting $V_{GL1}$, $V_{GL2}$, and $V_{GL3}$ to relatively high voltages. A single tunneling barrier is then formed in the left junction region near the lead electrode by varying $V_{GM1}$. Figs. 1(c,d) show the bias voltage spectra as a function of $V_{GM1}$ and vertical linecuts at specific gate voltage positions, revealing clearly a dual-gap feature. The larger gap remains fixed at ~ 1.3 meV and is unaffected by the middle-gate voltage, indicating that it originates from the parent Pb superconductor. In contrast, the smaller gap varies with the middle-gate bias; as the gate is made more positive, additional sub-gap states appear (see the positive-gate regime above 1 V in Fig. 1(c)), signaling the induced superconducting gap within the InSb nanowire. This behavior mirrors that already observed in Al-based hybrid systems, where gate tuning modifies the wave-function distribution across the SC-SM cross-section and consequently changes the SC-SM coupling strength [35,36,37]. Hence, in the relatively positive-gate regime the wavefunction is displaced away from the superconducting lead, weakening the coupling and reducing the induced gap. In the relatively negative-gate voltages, the wavefunction is close to the parent superconductor, resulting in strong coupling and the larger induced gap of ~1 meV. In particular in this regime, when the differential conductance is plotted on a logarithmic scale (as shown in Fig. 1(d)), a contrast ratio exceeding two orders of magnitude between the conductance outside and within the gap becomes evident, confirming the hardness of the induced gap [17-19]. This hard gap could be further proved by the zero nonlocal signal below the gap discussed later in Figs. 1(f,h). Also, the close size of parent and induced gap in the negative-gate regime here suggests strong SC-SM coupling, pointing to the native transparent hetero-interface in the hybrid device [25]. The same trend is reproduced in a second device (see Supplementary Fig. S1(b). We should emphasize that these dual-gap structure is not from the two gap nature of the bulk Pb because the inner one could be modulated by gate voltage as mentioned above, as well as it is accompanied by the switching positive-negative nonlocal signal in-between these two gaps which will be discussed later. The significantly larger and hard proximity-induced gap observed here, almost five times of that from Al-based configurations [17-19], implies potential better parameter space and larger topological gap protection for topological quantum states.

*These authors contributed to this work.
†Contact author: ziweidou@iphy.ac.cn
‡Contact author: e.p.a.m.bakkers@tue.nl
§Contact author: lilu@iphy.ac.cn
∥Contact author: shenjie@iphy.ac.cn

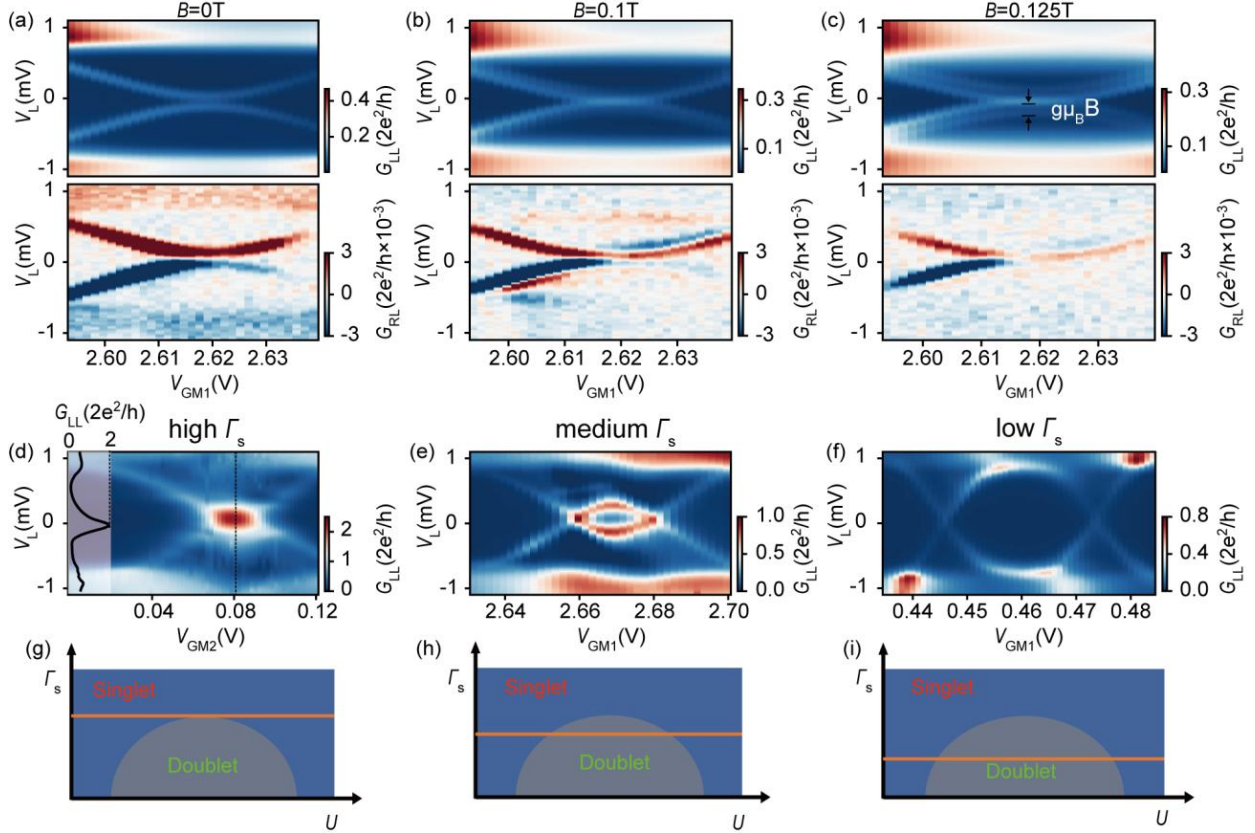

FIG. 2. Observation of the in-gap ABSs and its evolution with magnetic field and coupling strength between Pb and InSb. (a-c) $G_{LL}(V_L, V_{GM1})$ (upper panels) and $G_{RL}(V_L, V_{GM1})$ (lower panels) for out-of-plane magnetic field of $B = 0$ T, 0.1 T and 0.125T respectively, showing gate-tunable gapped ABS and its splitting due to the Zeeman effects. The estimated $g \approx 27.6$. (d-f) $G_{LL}(V_L, V_{GM2})$ (d) and $G_{LL}(V_L, V_{GM1})$ (e,f) for three typical couplings $\Gamma_S$ between Pb and InSb (see [33] for details). (g-i) The phase diagrams between the singlet and doublet states corresponding to (d-f), respectively. For high $\Gamma_S$ (d,g), the ABS is always in the singlet state, whose gap is closed with the appearance of a zero-bias peak (side-plot in (d)). For mediate to lower $\Gamma_S$ (e-f), the "eye-shape" corresponds the appearance of the doublet state.

*These authors contributed equally to this work.
†Contact author: ziweidou@iphy.ac.cn
‡Contact author: e.p.a.m.bakkers@tue.nl
§Contact author: lilu@iphy.ac.cn
∥Contact author: shenjie@iphy.ac.cn

The coexistence of the parent superconducting gap and the induced gap, as well as the hardness of the induced gap could be further confirmed by the nonlocal measurement in Figs. 1(e-i). The nonlocal signal exhibits zero-value below the smaller gap and non-zero values between the two gaps, confirming the distinct nature and origins of the two gaps by the following processes as illustrated in Fig. 1(j), as well as described in the previous Al-based devices [30,32,38,39]. When the applied bias voltage on the injection/the left lead is lower than the size of the induced gap in the hybrid segment, electrons in the normal electrode can only form Cooper pairs in the hybrid segment through Andreev reflection, as there is no quasiparticle state density within the hard induced gap, and subsequently enter the superconducting electrode. Consequently, no/zero nonlocal signal exists below the induced gap. This zero nonlocal conductance at bias voltage below the induced gap provides even more stronger evidence to confirm the hardness of the induced gap than the local spectrum. Similarly, when the bias voltage exceeds the gap of the parent superconductor, injected electrons can directly flow to ground via quasiparticle states above the gap in the superconducting electrode at the corresponding energy and are unlikely to generate nonlocal signals on the nonlocal side. Only when the bias voltage is between the induced gap and the parent superconducting gap, do electrons injected from one normal electrode reach the other (nonlocal) electrode, thereby generating a detectable positive/negative nonlocal signal (see Figs. 1(f,g), the reason will be discussed in the following paragraph) [32].

Moreover, a striking characteristic is a clear antisymmetry of the nonlocal signal [30,32,38,39] with respect to the bias voltage: As shown in Figs. 1(f,h), the nonlocal conductance takes a positive/negative value at positive bias voltages exceeding the induced gap and a negative/positive value at corresponding negative bias voltages. To analyze the antisymmetric nonlocal signal in Pb-InSb devices, as well as the switching positive-negative values, we varied the bias voltage $V_L$ to designate the left junction as the local contact, while maintaining $V_R$ at zero and treating the right junction as the nonlocal contact (see Fig. 1(j)). When varying the right gate voltage $V_{GM2}$ beneath the right junction, apparently it only changes the states in the right side because the left conductance (Fig. 1(e)) remained largely unchanged but the right conductance (Fig. 1(i)) varies. In this case, the nonlocal conductance in Fig. 1(f) was significantly modulated by $V_{GM2}$ which is the nonlocal gate and only tunes the right side, transitioning to an antisymmetric structure where the conductance became switching between negative and positive values across multiple gate voltages (in Fig. S2, another measurement configuration also confirms this). Through these comparisons, we conclude that the antisymmetric structure of the nonlocal signal in our Pb-InSb devices is primarily associated with the carrier in the junction on the nonlocal side. As illustrated in Fig. 1(j), as well as proved in previous Al-based devices [30,32,38,39], when the nonlocal carrier is electron-like, the nonlocal conductance at positive bias is positive (middle panel in Fig. 1(j)) and vice versa (right panel in Fig. 1(j)). (Here we define the nonlocal current flowing from the nonlocal normal lead to the grounding superconducting lead as positive, therefore the nonlocal signal at positive voltage is positive/negative when the charge component is electron/hole-like.) Similar characteristics were observed in all analogous aluminum-based three-terminal devices, where the overall sign reversal of the nonlocal conductance due to the nonlocal side's barrier was attributed to variations in tunnel barrier transmission with charge and energy, which is kind of random in this case [30,32,38,39].

*These authors contributed equally to this work.
†Contact author: ziweidou@iphy.ac.cn
‡Contact author: e.p.a.m.bakkers@tue.nl
§Contact author: lilu@iphy.ac.cn
∥Contact author: shenjie@iphy.ac.cn

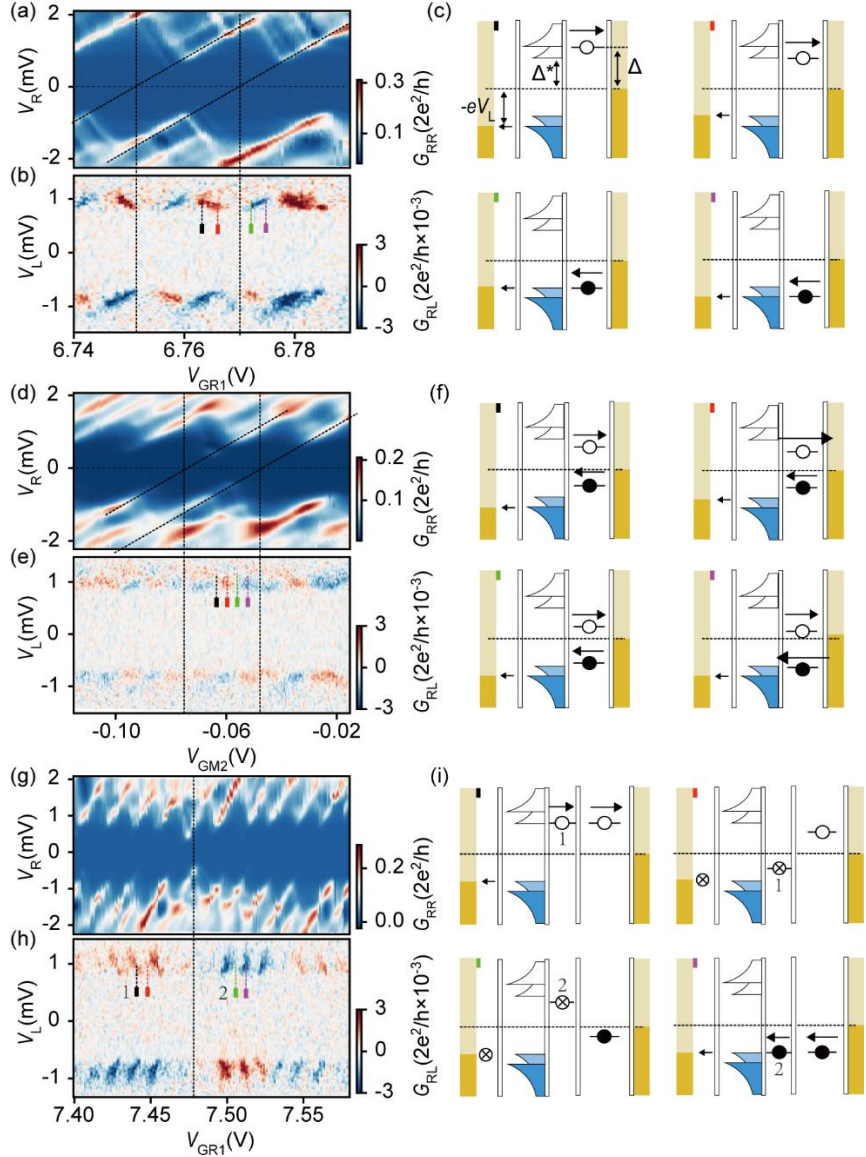

FIG. 3. Gate-modulated nonlocal conductance with different quantum dot configurations of the right junction: (a,b) $G_{RR}(V_R, V_{GR1})$ (a) and $G_{RL}(V_L, V_{GR1})$ (b). The local $G_{RR}(V_R, V_{GR1})$ (a) shows standard Coulomb oscillations, demonstrating the formation of the QD in the right junction ($V_L = 0$). (b) shows the nonlocal signals with alternating signs ($V_R = 0$). (c) The corresponding $G_{RL}$ processes marked in (b) with single QD level inside $\Delta^*$. The positions (marked by the vertical dashed lines) and periods of sign reversals match with that of $G_{RR}$, consistent with [32]. (d, e) Similar to (a,b) but measured in the second QD configuration ((d): $V_L = 0$, (e): $V_R = 0$). The positions (marked by the vertical dashed lines) and periods of sign reversals do not match with that of $G_{RR}$. (f) The corresponding $G_{RL}$ processes marked in (e), with two QD levels inside $\Delta^*$. (g,h) Similar to (a,b) but measured in the third QD configuration ((g): $V_L = 0$, (h): $V_R = 0$). $G_{RL}$ has three oscillations of the same sign and has the sign switch for roughly every three oscillations. (i) The corresponding $G_{RL}$ processes marked in (h), with two QDs in series (the right one is larger). All the arrows here point to the direction of electron. The full conductance matrix is shown in Fig S5.

*These authors contributed equally to this work.
†Contact author: ziweidou@iphy.ac.cn
‡Contact author: e.p.a.m.bakkers@tue.nl
§Contact author: lilu@iphy.ac.cn
∥Contact author: shenjie@iphy.ac.cn

Despite the fact that we observed the hard induced gap in the relatively negative back-gate voltages with the nearly zero nonlocal conductance, there are also significant subgap states when driving the gate voltages $V_{GM}$ to relatively positive values to push the electron wavefunction away from the parent superconductor [40]. Then, different types of ABSs could be visible. A typical gate-dependent ABS is shown in upper panel in Fig. 2(a), with positive/negative nonlocal signal indicating the electron/hole-like charge at positive/negative voltage bias, respectively (low panel in Fig. 2(a)). Once applying magnetic field, the spin-degeneracy ABS gradually becomes spin-split due to Zeeman field (Figs. 2(b,c), upper panels). The spin-split ABSs result in different sign of the nonlocal signal in Figs. 2(b,c) (lower panels), probably due to the spin-filter in another terminal [41]. We can extract a $g$-factor of $\sim$ 27.6 from the magnetic-field-dependent Zeeman energy $g\mu B$ marked in Fig. 2(c). With different gate voltages, we can tune the coupling between the resonance (or the unintentional QD) and the superconductivity ($\Gamma_S$), as well as modifying the charge energy ($U$) of the resonance (or the unintentional QD), and gradually vary the ABS from bent gap structure (Fig. 2(a)) to the closed gap with the zero-bias conductance peak (Fig. 2(d)), and then to eye shape (Figs. 2(e,f)), which indicates the singlet-to-doublet transition [42]. The zero-bias peak of ABS in Fig. 2(d) could reach the quantized value of $4e^2/h$ (left side of the upper panel). All these ABSs even in the positive gate regime, exhibit a significant $g$-factor with obvious spin-split energy at much lower magnetic field compared to that of Al-based hybrid devices (Fig. S4) [33]. Along with the different local ABSs, the nonlocal conductance presents both charge and spin filter signified by the reversal with respect to opposite bias and spin-split ABSs [41]. The extremely large gap along with a large variable energy for ABSs, great tunability of the wavefunction, and significant $g$-factor, confirmed by both local and nonlocal measurement, provide the fruitful phenomena to the Pb-based hybrid device and make it the better platform for topological qubits than Al-based devices.

To realize artificial Kitaev chain, spin-polarized quantum dot on the two sides have to form as shown in Fig. S3. After detailed characterizations of quantum dots on the nanowire using two finger gates to induce high barriers (see Sec. IV in [33]), we used the right junction region, where a quantum dot is formed, as the detector. In contrast, the left junction region, which serves only as an injector, retains a single tunnel barrier formed by $V_{GL1,2,3}$. The typical local and nonlocal measurements are shown in Figs. 3(a,b) respectively with a full conductance matrix in Fig.S5. Here $V_L$ is fixed at 0 such that Fig. 3(a) characterizes the local conductance of the QD on the right side with a hard induced gap illustrated also in Figs. S3(a,b,e) (the nonlocal conductance is shown in Fig. S5(c)). Since the horizontal dashed line corresponds to $V_R = 0$ mV, the intersection points between the horizontal and slanted dashed lines in Fig. 3(a) represent positions where the quantum dot energy levels align with the Fermi energy pointing to the charge degeneracy points, marked also by the vertical dashed lines in Fig. 3(a). On either side of these intersection points, the quantum dot energy levels allow relaxation only via electron or hole states, respectively. Fig. 3(b) shows the nonlocal conductance measured through the right detection junction as a function of the injection-junction voltage $V_L$, under the same $V_{GR1}$ range but with the fixed $V_R = 0$ (The local measurement of the left side is listed in Fig. S5(d), showing the dual-gap spectrum). Similar to Fig. 1(f), the nonlocal conductance exhibits a sign reversal depending on the gate voltage. In particular the gate-voltage interval of two such consecutive switches roughly matches the gate-voltage interval between the two charge degeneracy points of the quantum dot. One of the switching positions precisely coincides with the point where the quantum dot crosses the Fermi energy (indicated by the vertical dashed lines in Figs. 3(a,b)). This dual-antisymmetric behavior with respect to bias and gate voltages is associated with four distinct processes, depending on whether the nonlocal resonance is electron/hole-like, as schematically shown in Fig. 3(c). When applying the positive voltage bias to the left side driving the injection from the left normal lead, the nonlocal conductance is positive/negative if the corresponding nonlocal carrier on the right QD level from the middle superconducting lead to the right normal lead is electron/hole-like, as illustrated in the upper/lower panels. Therefore, by incorporating the quantum dot, we demonstrated that the antisymmetric nonlocal signals in our device arise from variations in the nonlocal tunneling probability with energy and charge, consistent with findings in Al-based devices [30,32]. Meanwhile, there is slight difference: In our case the nonlocal signal resonates as a function of both voltage bias and gate voltage instead of extending between the induced and parent gaps. This means only when the Fermi level of the injection (the left side) and the QD level of the detection (the right side) are aligned (see the lower panels in Fig. 3(c)), the carrier will contribute to the transport process and these are totally elastic relaxation [32]. It might also indicate that there is crossed Andreev reflection and elastic co-tunneling involved in the nonlocal process which results in positive and negative resonating nonlocal signal, respectively [11,26].

*These authors contributed equally to this work.
†Contact author: ziweidou@iphy.ac.cn
‡Contact author: e.p.a.m.bakkers@tue.nl
§Contact author: lilu@iphy.ac.cn
∥Contact author: shenjie@iphy.ac.cn

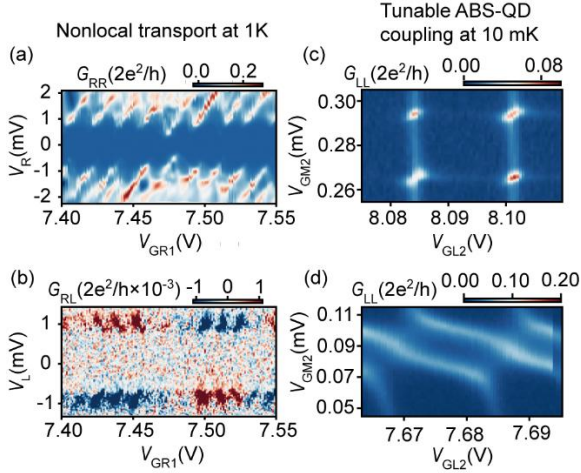

FIG. 4. Nonlocal transports at 1 K and charge stability diagrams showing tunable QDs and ABS coupling at 10 mK: (a, b) $G_{RR}(V_R, V_{GR1})$ (upper panels) and $G_{RL}(V_L, V_{R1})$ (lower panels) measured at 1 K, still showing nonlocal transport of the induced gap in the QD regime with the similar sign switch in $G_{RL}$ as Fig. 3(h). (c, d) Charge stability diagram $G_{LL}(V_{GM2}, V_{GL2})$ with $V_{L,R} = 0$ measured at 10 mK, in the weak coupling limit with the level crossings (c) and in the strong coupling limit with level anti-crossing (d).

Importantly, we observed distinct nonlocal conductance patterns in other quantum dot regimes, as shown in Figs. 3(d-f). While these patterns also exhibit dual antisymmetry with respect to voltage bias and gate, the positions and periods of sign reversals do not perfectly align with the charge degeneracy points of the quantum dot. We attribute this discrepancy to the large induced gap in Pb-based devices compared to the Al-based devices. As illustrated in Fig. 3(f), when the sum of the quantum dot's charging energy and level spacing is (twice) smaller than the induced gap, non-equilibrium relaxation involves two energy levels of the quantum dot in the detector. Consequently, the net current sign through the junction no longer depends solely on the relative position of a single quantum dot level and the Fermi energy, but also consider the energies and tunneling rates of both levels (symbolized by the relative arrow length near the quantum dots in the Fig. 3(f), the longer, the stronger). Simply speaking, if the two levels have equal rate, the net current switches the sign when the Fermi level locates at the center of two levels. In contrast, if the tunneling rates are unequal, the transition appears multiple times at random gate voltages within one level depending on the net carrier component, which is consistent with the observation in Figs. 3(d,e).

Besides the above-mentioned two cases, we observed the third which shows positive-negative switching crossing several levels instead within one level. Apparently, it switches at a special level with much big level spacing (pointed by dashed line in Figs. 3(g,h)), indicating a large QD connecting in series with a small one as shown in Fig. 3(i) (the rightmost one is the larger QD and the second-right one is the smaller QD). Interestingly, this additional level behaves like a charge filter, which only supports the process carried by the same charge in the two QDs as shown in the top-left and bottom-right panels, and forbids the process carried by different charges as shown in the top-right and bottom-left panels. Therefore, one type of nonlocal signal is always forbidden within the bigger QD level, e.g., the negative/positive value at the left/right side of the vertical line at positive bias in Figs. 3(g,h). Such a situation with two QDs connecting in series might happen due to using multiple finger gates to tune the chemical potential and induce a complex fluctuation.

Due to the high $T_c$ of Pb and the large induced gap, we warm up the device from ~ 10 mK to 1 K (Figs. 4(a,b)). Despite raising the temperature by two orders of magnitude, the hard induced gap in the QD diamond region and the positive-negative antisymmetric structure between the two gaps accompanied by the zero values below the induced gap in the nonlocal signal remain clearly visible. A larger proximity-induced superconducting gap directly translates into a proportionally larger topological gap. Because the topological gap sets the energy scale that protects Majorana modes, its increase makes the system far more resilient to disorder and charge fluctuations.

*These authors contributed equally to this work.
†Contact author: ziweidou@iphy.ac.cn
‡Contact author: e.p.a.m.bakkers@tue.nl
§Contact author: lilu@iphy.ac.cn
‖Contact author: shenjie@iphy.ac.cn

Consequently, one can extend the Kitaev chain to many more sites towards robust topological protection, paving the way toward scalable networks of topological qubits. This high temperature approaches the critical temperature of parent Al and much higher than that of the Al-based Kitaev chain, demonstrating that Pb-based devices exhibit superior stability against thermal fluctuations.

To perform Kitaev chain, the QDs and ABS coupling should also be present and tunable [11-15,43]. We illustrate the two extreme cases in Figs. 4(c,d) taken in 10 mK, where crossings shows up in the charge stability diagram for the weak coupling limit and anti-crossings for the strong coupling limit. Here, we can only found this anti-crossings which is similar to the ECT-like process, but not the CAR-like. The reason might be there is always charging energy for ABSs which is not gapped but eye-shape like. For Kitaev chain realization, once the grounding superconducting leads with ABSs replaced by the floated ones or the ABSs acquire charging energy, it is also possible to fine-tune the gate to reach a sweet point [44,45]. Of course, the charging energy could also be eliminated by stronger SC-SM coupling, which is reachable by increasing the etching time [24] and applying negative gate voltage to push the wavefuction towards the parent superconductor [35-37].

In summary, in the Pb-InSb three terminal device, we harness the nonlocal measurement to reveal the dual-gap signature with an extremely large and hard induced gap, because a finite non-local conductance appears only for bias voltages that lie between the induced and parent gaps. This provides an unambiguous probe of the dual-gap structure that cannot be accessed with conventional two-terminal spectroscopy. In particular, antisymmetric bias dependence in-between shows the polarity switches between positive- and negative- values, pointing to electron- and hole-like carrier. The gap is also tunable with different ABSs signaling singlet-to-doublet transition. Furthermore with the assistance of spin-polarized QD levels, the non-local signal switches sign when the QD level crosses the Fermi energy, and multiple resonances produce richer antisymmetric structures, illustrating charge- and spin-filtering effects. These nonlocal signatures persist up to ~ 1 K accompanied by the unusually large induced gap (~1 meV), far exceeding both the operating temperature (~10 mK) and the induced gap (~0.2 meV) of Al-based devices. This demonstrates the superior robustness to perturbations of Pb-based hybrid devices. Once coupling the ABSs with spinful QDs, anti-crossing and crossing show up, revealing the different coupling strength by fine-tuning gate voltage and paving the way towards Kitaev chains. Overall, the nonlocal transport data provide decisive evidence for a hard, large, tunable induced gap, and the ability to manipulate charge and spin transport through quantum-dot resonances – all the essential ingredients for realizing artificial Kitaev chains and topological superconductivity in Pb-InSb platforms, in particular potentially with a substantially larger topological gap that facilitates future scaling.

We are grateful to Chunxiao Liu, Leo Kouwenhoven, Nick van Loo, Francesco Zatelli, Srijit Goswami Micheal Wimmer and Michele Burrello for helpful discussions. The work of Z.D. and J.S. was supported by the Beijing Natural Science Foundation (Grant No. JQ23022), the Young Scientists Fund of the National Natural Science Foundation of China (Grant No. 2024YFA1613200) and the National Key Research and Development Program of China (Grant Nos. 2023YFA1607400). The work of Z.D. was supported by the National Natural Science Foundation of China (Grant No. 12504561). The work of D.P. and J.Z. was supported by the National Natural Science Foundation of China (Grant Nos. 12374459, 61974138 and 92065106), the Innovation Program for Quantum Science and Technology (Grant 2021ZD0302400). D. P. acknowledges the support from Youth Innovation Promotion Association, Chinese Academy of Sciences (Nos. 2017156 and Y2021043). The work of J.S., L.L., F.Q. and G.L. were supported by the National Natural Science Foundation of China (Grant Nos. 12174430, 92365302), and the Synergetic Extreme Condition User Facility (SECUF, https://cstr.cn/31123.02.SECUF). Y.L. acknowledges support from National Natural Science Foundation of China, Grant No. 12404154. The work of other authors were supported by the National Key Research and Development Program of China (Grant Nos. 2019YFA0308000, 2022YFA1403800, 2023YFA1406500, and 2024YFA1408400), the National Natural Science Foundation of China (Grant Nos. 12274436, 12274459), the Beijing Natural Science Foundation (Grant No. Z200005), and the Synergetic Extreme Condition User Facility (SECUF, https://cstr.cn/31123.02.SECUF). The work is also funded by Chinese Academy of Sciences President's International Fellowship Initiative (Grant No. 2024PG0003). MPN acknowledges support by the National Science Center, Poland (NCN) agreement number UMO-2020/38/E/ST3/00418.

*These authors contributed equally to this work.
†Contact author: ziweidou@iphy.ac.cn
‡Contact author: e.p.a.m.bakkers@tue.nl
§Contact author: lilu@iphy.ac.cn
‖Contact author: shenjie@iphy.ac.cn

*These authors contributed to this work.
†Contact author: ziweidou@iphy.ac.cn
‡Contact author: e.p.a.m.bakkers@tue.nl
§Contact author: lilu@iphy.ac.cn
‖Contact author: shenjie@iphy.ac.cn

*These authors contributed equally to this work.
†Contact author: ziweidou@iphy.ac.cn
‡Contact author: e.p.a.m.bakkers@tue.nl
§Contact author: lilu@iphy.ac.cn
‖Contact author: shenjie@iphy.ac.cn

*These authors contributed equally to this work.
†Contact author: ziweidou@iphy.ac.cn
‡Contact author: e.p.a.m.bakkers@tue.nl
§Contact author: lilu@iphy.ac.cn
∥Contact author: shenjie@iphy.ac.cn

# Supplemental Material: Broad nonlocal spectrum in the Pb-InSb hybrid three terminals for potential realization of Kitaev chains

## I. Sample fabrication and measurement technique

*Sample fabrication*: The thin finger backgates are first fabricated on a silicon substrate by high-definition electron beam lithography, followed by the deposition of 3/7 nm Ti/Au. The exfoliated hexagon boron nitride (hBN) is then placed unto the gate area as the dielectric layer. With the aid of a micro-manipulator tip under the optical microscope, high mobility InSb nanowires grown by metal organic vapor phase epitaxy (MOVPE) [1] are then transferred from the growth chip to the substrate covered by hBN. The 5/105 nm Ti/Au electrodes are subsequently fabricated by the standard electron beam process. As the final step, 110 nm Pb in the middle terminal is defined and deposited. In order to obtain good contact and appropriate band bending at the hetero-interface, we use gentle argon milling to remove the oxide layer of the nanowire and induce an smooth and accumulation layer on the surface before each deposition, as proved by our previous devices [2]. The uncovered section of the nanowire between the Ti/Au normal metal electrode and the Pb superconductor electrode serves as a tunnel barrier, which is tunable by the electrostatic gate. Immediately after depositing the Pb shells, the aluminum oxide ($AlO_x$) is evaporated to prevent the oxidation and dewetting of the Pb.

*Full conductance matrix measurement*: We adopt the widely used methods [3,4,5,6] to measure the full conductance matrix defined in Eq. (1) in the main text in our three-terminal device. In order to differentiate the nonlocal current from the local signals, two lock-in amplifiers provide the ac currents $dI_L$ and $dI_R$ with the different frequencies (~7.3 Hz and ~13.2 Hz) to the left and right Ti/Au electrodes, and also measure the local differential conductance $G_{LL} = dI_L/dV_L$ and $G_{RR} = dI_R/dV_R$ of the respective junctions. The nonlocal differential conductance $G_{LR} = dI_L/dV_R$ and $G_{RL} = dI_R/dV_L$ are measured with two additional lock-in amplifiers synchronized with the driving lock-in at the opposite junction. The two different frequencies thus enable the simultaneous readout of the full conductance matrix. The 2D plots are then measured with the same parameters but swept twice with the dc bias voltage $V_R$ and $V_L$, respectively.

## II. Gate-tunable induced gap in Device 2

Fig. S1(a) shows the similar Device 2. At the fixed $V_{GM} = 0$ V, the local measurement of the left tunnel junction conductance $G_{LL}$ versus $V_L$ and $V_{GL3}$, showing the gap of Pb $\Delta = 1.2$ meV and the induced gap $\Delta^*$, similar to Fig. 1. Using a different $V_{GM} = 1.8$ V, the induced gap varies, demonstrating its strong gate tunability.

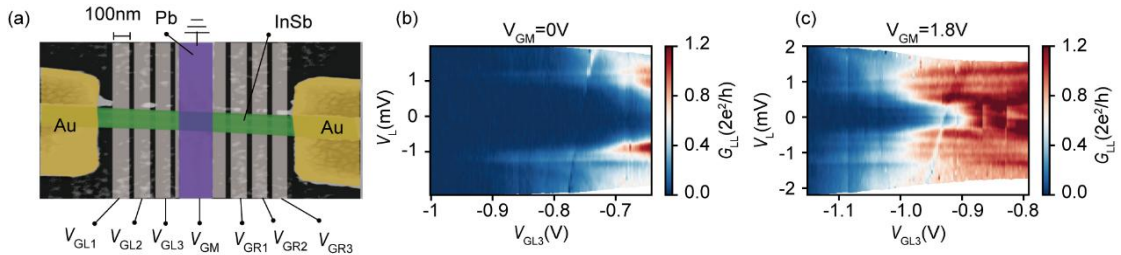

Fig. S1. Gate-tunable induced gap measured in a similar Device 2: (a) SEM of Device 2 (b) $G_{LL}$ versus $V_L$ and $V_{GL3}$, showing the gap of Pb $\Delta$ and the similar induced gap $\Delta^*$. $V_{GM} = 0$ V (c) The induced gap is measured with a different $V_{GM} = 1.8$ V.

### III. Additional local and nonlocal data measured in Device 1

Figs. S2(a-c) show the local and nonlocal conductance dependence on $V_{GM1}$, which controls the left junction. The nonlocal $G_{RL}$ is hardly modulated by $V_{GM1}$. Meanwhile, similar measurements with $V_{GM2}$ (Figs. S2(d-f)) show a strong modulation of $G_{RL}$ by $V_{GM2}$, which controls the right junction/the nonlocal junction. Combining the two measurements, the variation of nonlocal signal indeed originates from the right junction/nonlocal junction.

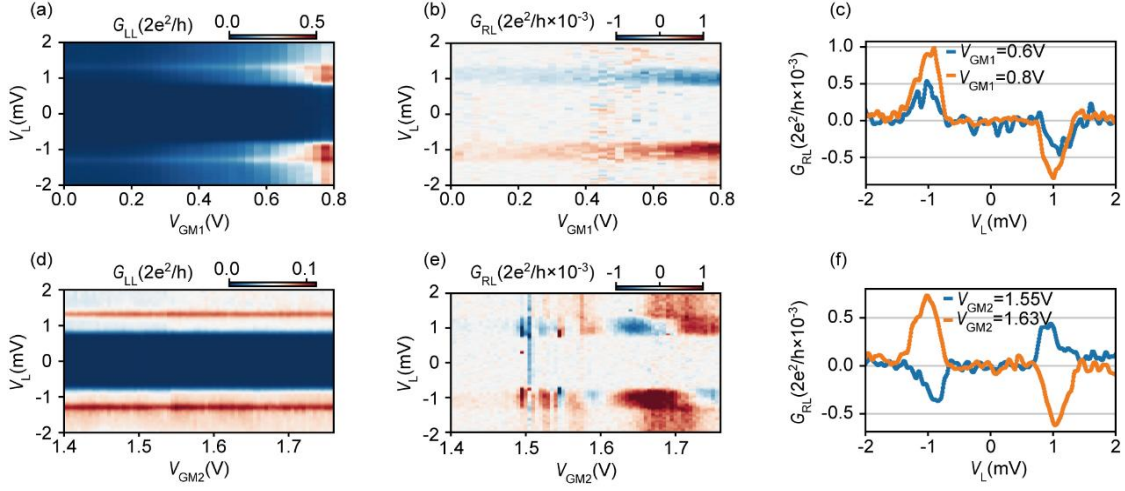

Fig. S2: Additional local and nonlocal data measured in Device 1: (a-c) local and nonlocal conductance dependence on $G_{GM1}$: $G_{LL}(V_L, V_{GM1})$, $G_{RL}(V_L, V_{GM1})$ and the linecuts $G_{RL}(V_L)$ in (b), respectively. (d-e) local and nonlocal conductance dependence on $G_{GM2}$: $G_{LL}(V_L, V_{GM2})$, $G_{RL}(V_L, V_{GM2})$ and the linecuts $G_{RL}(V_L)$ in (e), respectively.

### IV. Formation of the QDs in both right and left junctions

Figs. S3(a, b) show the local conductance of the right and the left junctions $G_{RR}$ and $G_{LL}$ respectively, with both the hard induced gap (horizontal lines) and the "Coulomb diamonds" (slanted lines) typical of the QDs. The diamonds present even-odd alternating structure because of the co-existence of both level spacing and charge energy, so it is spinful. Fig. S3(c) shows the evolution of Coulomb resonance peaks $G_{RR}(V_{GR1})$ at zero voltage bias with out-of-plane magnetic field $B$, revealing the modulation of the peak positions with $B$ and the associated level shift in opposite directions due to the spin-polarized QD levels [2,7,8]. Meanwhile, by fixing the gate voltage off resonance, the bias spectroscopy $G(V_R)$ with $B$ in Fig. S3(d) shows the splitting of single QD levels with two spin-polarized states by the Zeeman energy [2,7,8]. In particular, at the finite $B$, a pair of spin-up and spin-down levels crosses and an anti-crossing happens due to SOC [9]. In addition to Figs. S3(a-d) measured on Device 1, the similar QD Coulomb diamonds are also reproduced in Device 2, with the clear dual-gap structure indicating the parent gap and the induced gap.

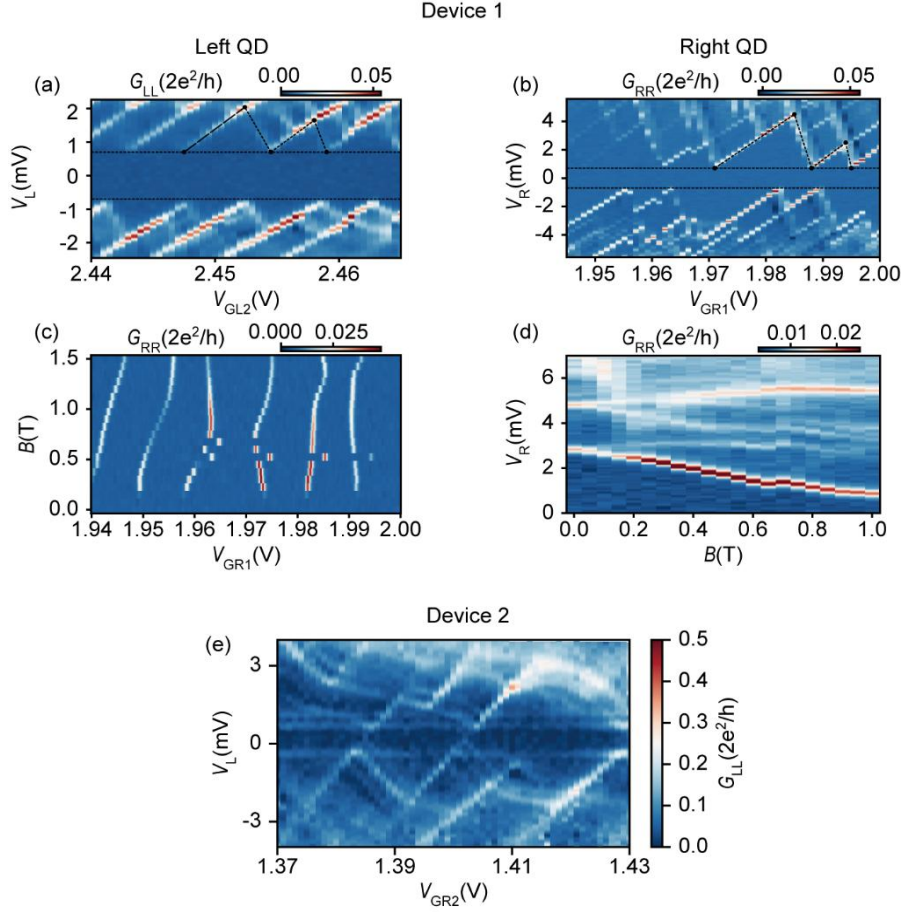

Fig. S3. Characterization of the QDs in the right and the left junctions: (a) $G_{LL}(G_{GL2}, V_L)$ at $B = 0$ for left QD. (b) $G_{RR}(V_{GR1}, V_R)$ at $B = 0$ for right QD. Both (a, b) show the induced gap (horizontal lines) and the Coulomb diamonds (slanted lines). (c) $G_{RR}(V_{GR1}, B)$ at $V_R = 1$ V. (d) $G_{RR}(V_R, B)$ at $V_{GR1} = 2$ V, showing spin splitting of the QD levels. (a-d) are measured in Device 1. (e) Additional QD measurements on Device 2 showing the clear induced gap.

## V. Additional data of the in-gap ABSs and its evolution with magnetic field

Fig. S4 shows the in-gap ABSs with the characteristic "eye-shape", indicating the medium-high coupling regime as explained in Fig. 2. Their further evolution with magnetic field shows the Zeeman splitting of the levels, from which we can estimate the g-factor to be ~ 27.6.

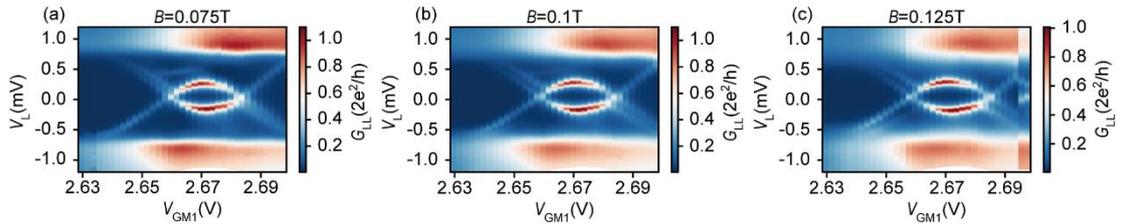

Fig. S4. Additional data of the in-gap ABSs and its evolution with magnetic field (Device 1): (a-c) $G_{LL}(V_L, V_{GM1})$ at $B = 0.075$ T, $B = 0.1$ T, and $B = 0.125$ T.

## VI. Full conductance matrix data for Fig. 3.

Fig. S5 shows the full conductance matrix data for the three cases in Fig. 3, with $G_{RR}$ and $G_{RL}$ reproduced from Fig. 3. The comparison between $G_{RR}$ and $G_{LL}$ confirms that the QDs are located in the right junction in each case, while a dual-gap structure is present in the left side.

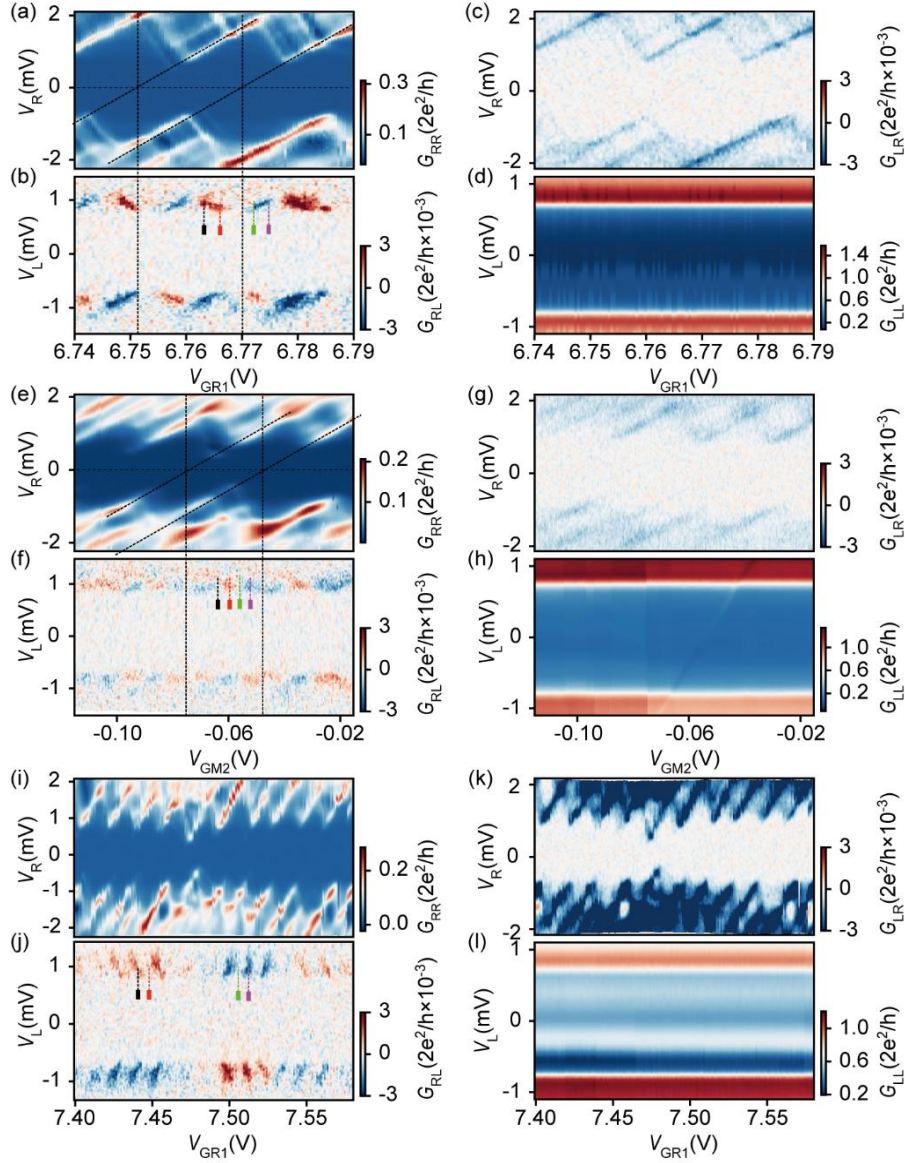

Fig. S5. Full conductance matrix data for Fig. 3 for the three QD configurations (Device 1): (a-d) $G_{RR}(V_R, V_{GR1})$, $G_{RL}(V_L, V_{GR1})$, $G_{LL}(V_L, V_{GR1})$, $G_{LR}(V_R, V_{GR1})$ for the first configuration, respectively. (a,b) are reproduced from Figs. 3(a,b). (e-h) $G_{RR}(V_R, V_{GM2})$, $G_{RL}(V_L, V_{GM2})$, $G_{LL}(V_L, V_{GM2})$, $G_{LR}(V_R, V_{GM2})$ for the second configuration, respectively. (g,h) are reproduced from Figs. 3(d,e). (i-l) $G_{RR}(V_R, V_{GR1})$, $G_{RL}(V_L, V_{GR1})$, $G_{LL}(V_L, V_{GR1})$, $G_{LR}(V_R, V_{GR1})$ for the third configuration, respectively. (i,j) are reproduced from Figs. 3(g,h).

## VII. Additional data for nonlocal measurement at 1 K and the stability diagram measurement at 10 mK

Figs. S6(a, b) show the QD Coulomb diamond in the local signal and the corresponding nonlocal signals at 1 K as Figs. 4(a,b), with the QD configuration similar to Fig. 3(c). Fig. S6(c) shows the local tunnel spectroscopy measurement of $G_{LL}$ where an accidental ZBP appears, in the range of $B$ outside the critical field of the induced superconductivity. This is possibly due to the QD level which is on resonance with the electrode. Usually, the QD levels are modulated with $B$, as observed in Fig. S3(c). Here, the stable ZBP with $B$ possibly originates from the location of ($V_L$, $B$) where the anti-crossing of the spin-polarized QD levels happens with the SOC [2,7,8,10]. This makes the level less sensitive to $B$ and may produce the stable ZBP here extending for ~ 0.2 T. We note that the MZMs cannot cause such signal as it appears outside the critical field. Fig. S6(d) is the stability diagram measured at 10 mK at a different gate configuration.

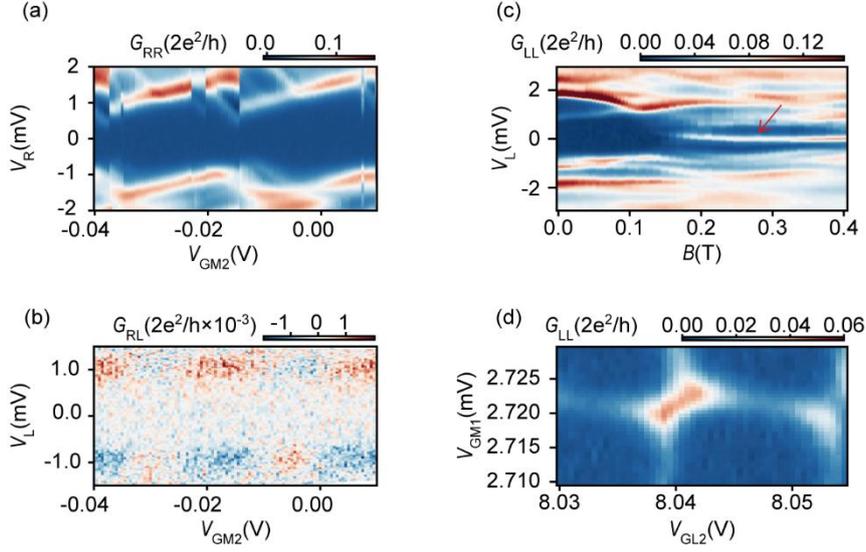

Fig. S6. (a, b) $G_{RR}(V_R, V_{GM2})$ and $G_{RL}(V_L, V_{GM2})$ measured at 1 K under different gate configurations as Figs. 4(a, b), again showing nonlocal transport of the induced gap in the QD regime with the similar sign switch in $G_{RL}$ as Fig. 3(b). (a, b) are measured at 1 K. (c) Accidental zero-bias-peak (ZBP, marked by a red arrow) appeared outside the critical field of the device, due to the QD level on resonance with the electrode. (d) Stability diagram with another gate configuration. (c,d) are measured at 10 mK.